\documentclass[aps,prd,twocolumn,superscriptaddress,showpacs,amsmath,amssymb,amsthm,nofootinbib, preprintnumbers]{revtex4-2}
 \usepackage{amsmath}
\usepackage{graphicx}
\usepackage{epstopdf}
\usepackage{float}
\usepackage{hyperref}
\usepackage{color}
\usepackage[T1]{fontenc}
\usepackage[utf8]{inputenc}
\usepackage[toc,page]{appendix}
\usepackage[usenames,dvipsnames]{xcolor}
\usepackage[normalem]{ulem}

\newcommand{\be}{\begin{equation}}
\newcommand{\ee}{\end{equation}}
\newcommand{\ba}{\begin{eqnarray}}
\newcommand{\ea}{\end{eqnarray}}

\newcommand{\beq}{\begin{equation}}
\newcommand{\eeq}{\end{equation}}
\newcommand{\beqa}{\begin{eqnarray}}
\newcommand{\eeqa}{\end{eqnarray}}
\newcommand{\nn}{\nonumber}

\begin{document}

\title{Slowly rotating black holes in nonlinear electrodynamics}

\author{David Kubiz\v n\'ak}
\email{dkubiznak@perimeterinstitute.ca}
\affiliation{Institute of Theoretical Physics, Faculty of Mathematics and Physics,
Charles University, Prague, V Hole{\v s}ovi{\v c}k{\' a}ch 2, 180 00 Prague 8, Czech Republic}
\affiliation{Perimeter Institute, 31 Caroline Street North, Waterloo, ON, N2L 2Y5, Canada}

\author{Tayebeh Tahamtan}
\email{tahamtan@utf.mff.cuni.cz}
\affiliation{Institute of Theoretical Physics, Faculty of Mathematics and Physics,
Charles University, Prague, V Hole{\v s}ovi{\v c}k{\' a}ch 2, 180 00 Prague 8, Czech Republic}

\author{Otakar Sv{\'i}tek}
\email{ota@matfyz.cz}
\affiliation{Institute of Theoretical Physics, Faculty of Mathematics and Physics,
Charles University, Prague, V Hole{\v s}ovi{\v c}k{\' a}ch 2, 180 00 Prague 8, Czech Republic}


\date{March 3, 2022}

\begin{abstract}
We show how (at least in principle) one can  construct electrically and magnetically charged slowly rotating black hole solutions coupled to non-linear electrodynamics (NLE). Our generalized Lense--Thirring ansatz is, apart from the static metric function $f$ and the electrostatic potential $\phi$ inherited from the corresponding spherical solution, characterized by two new functions $h$ (in the metric) and $\omega$ (in the vector potential) encoding the effect of rotation. In the linear Maxwell case, the rotating solutions are completely characterized by static solution, featuring $h=(f-1)/r^2$ and $\omega=1$. We show that when the first is imposed, the ansatz is inconsistent {with any restricted (see below) NLE but the Maxwell electrodynamics.} In particular, this implies that the {(standard) Newman--Janis} algorithm cannot be used to generate rotating solutions for any restricted non-trivial NLE. {We present a few explicit examples of slowly rotating solutions in  particular models of NLE, as well as briefly discuss the NLE charged Taub-NUT spacetimes. } 
\end{abstract}

\maketitle

\section{Introduction}

{Theories of {\em non-linear electrodynamics} (NLE) are classical field theories that naturally generalize the linear Maxwell theory. 
Dating back to the beginning of the twentieth century, first such theories   
emerged as an attempt to tame the divergencies associated with point-like charges and cure the problem of infinite self-energy in Maxwell's theory. While this original problem has later been resolved by invention of renormalization, NLE remains at the theoretical forefront  to these days. Perhaps the best known example of NLE is the Born--Infeld theory \cite{BornInfeld}, which has many unique and remarkable properties \cite{Plebanski:1970zz}, and naturally arises in the context of string theory \cite{fradkin1985non, Polchinski:1996na} and early universe cosmology \cite{Alishahiha:2004eh}. Other models of NLE were proposed to resolve the spacetime singularity \cite{Ayon-Beato:1998hmi} -- providing a physical source for the regular black holes \cite{bardeen1968non}, to capture the basic features of the QED at the classical level \cite{Heisenberg:1936nmg}, to describe dual strings in flat spacetime \cite{Nielsen1973}, or most recently as a maximally symmetric alternative to the Maxwell theory \cite{Bandos:2020jsw, Kosyakov:2020wxv} and its deformation \cite{Babaei-Aghbolagh:2022uij}.
}

Of course, all the above models can be straightforwardly extended to a curved spacetime. At present there exists a plethora of static {\em spherically symmetric} solutions of Einstein-NLE system -- such solutions are known for the Born--Infeld theory \cite{Bicak:1975dq, Demianski:1986wx,Fernando:2003tz}, for logarithmic Lagrangians \cite{Soleng:1995kn}, for a square root Lagrangian \cite{Aurilia:1993qi,Guendelman:1997mc}, for regular black hole models \cite{Ayon-Beato:1998hmi,Ayon-Beato:2000mjt,Bronnikov:2000vy} (see also \cite{Cano:2020ezi, Cano:2020qhy} for non-minimal coupling models) and other theories \cite{deOliveira:1994in,Maeda:2008ha, Tahamtan-fR-MAX:2012, Alvarez:2022upr}. Recently, also dynamical solutions lacking any symmetry were constructed \cite{Tahamtan-NE:2016} and their further generalization including electromagnetic radiation were studied in \cite{Tahamtan-PRD:2021, Kokoska:2021lrn}. All these solutions are, however, twist-free and it would be extremely valuable to obtain rotating generalization of the static spherically symmetric cases thus providing NLE version of the 
Kerr--Newman solution.

{So far there have been a number of attempts at constructing {\em rotating} black holes coupled to NLE. An obvious candidate to this end is to try to generate such solutions from the corresponding static ones by employing the (possibly upgraded) {\em Newman--Janis trick} \cite{Newman:1965tw, Azreg-Ainou:2014pra}. While this trick successfully leads to Kerr--Newman solution, it does not preserve the Einstein field equations for arbitrary source \cite{Drake:1998gf}, nor it works for the vacuum solutions in the presence of modified gravity theories \cite{Hansen:2013owa, Gray:2021roq}. In particular, charged spacetimes generated in this way, e.g. \cite{CiriloLombardo:2004qw, Dymnikova:2006wn, Toshmatov:2014nya, Dymnikova:2015hka,Toshmatov:2017zpr}, do not satisfy the corresponding Einstein-NLE equations \cite{CiriloLombardo:2004qw, Rodrigues:2017tfm, Breton:2019arv}.
At the same time, a task of solving the corresponding equations of motion directly seems, due to their innate non-linearity, quite formidable.
} That is why here we approach the rotating generalization more modestly -- by considering a {\em slow rotation approximation}.\footnote{{While some of the studies claim to have considered slowly rotating solutions in NLE, e.g. \cite{CiriloLombardo:2005qmc, Hendi:2010bk, Hendi:2014xia, Lammerzahl:2018zvb}, as we shall see below, this is not really the case because of the incorrect ansatz for such solutions.}} 

{As we shall see, Maxwell's theory is surprisingly unique in providing (even slowly) rotating solutions in a `natural way'. Specifically, we will prove a ``No Go theorem'' that in particular rules out the standard Newman--Janis trick as a way of deriving rotating NLE solutions within a large family of NLE models.
Based on this theorem, one can immediately dismiss a number of attempts at constructing rotating NLE black holes published previously in the literature.  
We will also present a complete set of equations governing slowly rotating Einstein-NLE solutions in a broader setting and use them to derive two new explicit solutions.
These solutions may, at least in principle, be used in future as `test grounds' for finding a possible generalization of the Newmann--Janis trick for the NLE theories (if it exists). 
}

In order to construct slowly rotating solutions in NLE we employ the  {
{\em generalized Lense--Thirring} ansatz for the metric  \cite{lense1918influence, Gray:2021toe, Gray:2021roq}:
\be\label{metric}
ds^2=-Nfdt^2+\frac{dr^2}{f}+2ar^2h\sin^2\!\theta dtd\varphi+r^2d\Omega^2\,,
\ee
where $a$ is the rotation parameter, $N=N(r)$ and $f=f(r)$ are two independent metric functions, and $d\Omega^2=\sin^2\!\theta d\varphi^2+d\theta^2\,$
is the volume element on the sphere. 
As we shall see, for any NLE one can set 
\be\label{N}
N=1\,, 
\ee 
generalizing the result of \cite{Jacobson:2007tj} for spherical Maxwell and Born--Infeld black holes. Moreover, for the charged slowly rotating solutions in the Einstein--Maxwell theory we have
\be\label{h}
h=\frac{f-1}{r^2}\,, 
\ee
where $f$ is the corresponding static metric function, $f=1-\frac{2M}{r}+\frac{e^2+p^2}{r^2}$, where $e,p$ are the electric, magnetic charges and $M$ stands for the mass. Interestingly, as we shall see in Sec.~\ref{Sec3}, the metrics generated by the Newman--Janis algorithm are, in the slow rotation approximation, of the form \eqref{metric} with \eqref{N} and \eqref{h}. However 
such a form, namely that $h$ is given by the corresponding static metric function $f$ via \eqref{h}, is  consistent {\em only} in the Maxwell theory among all NLEs of the restricted form \eqref{restrict} below. This in particular means that not only is the Newman--Janis algorithm unable to construct full rotating solutions in NLE, it actually fails already at the lowest linear in $a$ order. 
}

Our paper is organized as follows. In the next section we summarize the basics of NLE theories and list their equations. In Sec.~\ref{Sec3} we review the corresponding static solutions, as well as review the rotating metrics generated by the Newman--Janis formalism and their slowly rotating approximation. We then show how (at least in principle) one can construct electrically (Sec.~\ref{Sec4a}) and magnetically (Sec.~\ref{Sec4b}) charged solutions in any NLE, as well as establish the uniqueness of the Maxwell theory as the only NLE whose slowly rotating solutions can be written in the above form.  We conclude in Sec.~\ref{Sec5}. 
In Appendix~\ref{AppB} we construct slowly rotating magnetized solutions in the ``Square Root'' model of NLE,
Appendix~\ref{AppA} contains the discussion of NLE charged Taub-NUT solutions.

\section{Theories of NLE}\label{Sec2}
{Let us first review the basics of non-linear electrodynamics. Any such theory is formulated in terms of the two invariants of the electromagnetic field:} 
\be\label{SP}
{\cal S}=\frac{1}{2}F_{\mu\nu} F^{\mu\nu}\,,\quad {\cal P}=\frac{1}{2}F_{\mu\nu} (*F)^{\mu\nu}\,, 
\ee
where the field strength $F_{\mu\nu}$ is given in terms of the 
vector potential $A_\mu$ by the familiar expression, 
$F_{\mu\nu}=\partial_\mu A_\nu-\partial_\nu A_\mu$. {Whereas 
${\cal S}$ is a true scalar, the invariant ${\cal P}$ is only a pseudoscalar. To restore parity invariance, we thus consider theories that depend on ${\cal P}$ via its `square', that is, we assume that the NLE theory is characterized by the following Lagrangian: 
\be\label{general}
{\cal L}={\cal L}({\cal S},{\cal P}^2)\,. 
\ee
As we shall see, the latter assumption significantly simplifies the subsequent  discussion of the slowly rotating solutions. In addition, one might require that the theory of NLE should approach that of Maxwell in the weak field approximation, imposing  
\be\label{POC}
\lim_{F_{\mu\nu}\to 0} {\cal L}=\frac{1}{2}{\cal S}+O({\cal S}^2, {\cal P}^2)\,,
\ee
which is known as the {\em principle of correspondence}. (This requirement is violated, for example, by the Square Root Lagrangian discussed in the appendix.) 
}

{Introducing the following notation:
\be
{\cal L}_{\cal S}=\frac{\partial {\cal L}}{\partial {\cal S}}\,,\quad  
{\cal L}_{\cal P}=\frac{\partial {\cal L}}{\partial {\cal P}}=2{\cal P}{\cal L}_{{\cal P}^2}=2{\cal P}\frac{\partial {\cal L}}{\partial {\cal P}^2}\,, 
\ee
the {\em generalized Maxwell} equations read}
\be\label{FE}
d*E=0\,,\quad
dF=0\,, 
\ee
where 
\be
E_{\mu\nu} = \frac{\partial \mathcal{L}}{\partial F^{\mu\nu}}
=2\Bigl({\cal L_S}F_{\mu\nu}+{\cal L_P}*\!F_{\mu\nu}\Bigr)\,. 
\ee
Moreover, upon minimally coupling to the Einstein--Hilbert {term,
\begin{equation}\label{bulkAct}
    I= \frac{1}{16\pi} \int_{M} d^4x \sqrt{-g}\left(R -4{\cal L}\right)\,,
\end{equation}
we obtain the following 
{\em Einstein equations}: 
\be \label{Hmunu}
H_{\mu\nu}=G_{\mu\nu}-8\pi T_{\mu\nu}=0\,,
\ee
 where the generalized EM energy-momentum tensor reads 
\be\label{Tmunu}
T^{\mu\nu}=\frac{1}{4\pi}\Bigl(2F^{\mu\sigma}F^{\nu}{}_\sigma {\cal L_S}+{\cal P}{\cal L_P} g^{\mu\nu}-{\cal L}g^{\mu\nu}\Bigr)\,,
\ee
{We refer to equations \eqref{FE} and \eqref{Hmunu} as the Einstein-NLE equations.}  

{
In what follows, 
we shall also consider a }
simpler class of theories,  obtained by considering {\em restricted Lagrangians} that are independent of the invariant ${\cal P}$, that is
\be\label{restrict}
{\cal L}={\cal L}({\cal S})\,. 
\ee
The corresponding equations of motion straightforwardly follow from the above.

\section{Newman--Janis algorithm:
Generating rotating solutions from the static ones?}\label{Sec3}
{
In the NLE literature, many rotating black hole `solutions' have been  generated from the static ones by applying the {(standard -- `coined for Maxwell's theory')} Newman--Janis algorithm. In this section we review this approach and its limitations. We start by considering the static spherically symmetric solutions.   
}

\subsection{Static solutions}

{
Consider a general static spherically symmetric metric element,
\be\label{sss}
ds^2=-Nfdt^2+\frac{dr^2}{f}+r^2d\Omega^2\,,
\ee
where $N=N(r)$ and $f=f(r)$ are two independent metric functions, and $d\Omega^2=\sin^2\!\theta d\varphi^2+d\theta^2\,$
is the volume element on the sphere. It was shown in 
\cite{Jacobson:2007tj} for the Born--Infeld theory, and is similarly valid for any NLE, that 
\be\label{N1}
N=1\,. 
\ee 
The argument goes as follows. Let $l$ denotes a radial null vector of the metric element \eqref{sss}. Then we have $R_{\mu\nu}l^\mu l^\nu\propto N'\,.$
Thus, Eq.~\eqref{N1} can be imposed provided 
\be\label{Tll}
T_{\mu\nu}l^\mu l^\nu=0\,. 
\ee
The NLE electromagnetic stress tensor $T_{\mu\nu}$, \eqref{Tmunu} consists of two terms, first proportional to $g_{\mu\nu}$ and the second proportional to $F_{\mu\alpha}F_\nu{}^{\alpha}$\,. When multiplied by $l^\mu l^\nu$, the first term trivially vanishes, while the latter term is proportional to $w_\alpha w^\alpha$, where $w_\alpha=F_{\alpha\mu}l^\mu$. However, for radial magnetic fields we have $w_\alpha=0$, whereas for radial electric fields $w^2=0$, implying that \eqref{Tll} is satisfied for any NLE, and \eqref{N1} can be imposed \cite{Jacobson:2007tj}. 
}

{
To find the spherical solution for given NLE,  we thus consider the spherical element \eqref{sss} with $N=1$, supplemented by the corresponding vector potential.  
Considering both electric $e$ and magnetic $p$ charges, the ansatz for the vector potential reads
\be
A=e\phi dt+p \cos\theta d\varphi\,,
\ee
where $\phi=\phi(r)$ is the function characterizing the electrostatic potential. 
We then find that the invariants \eqref{SP} are given by 
\be
 {\cal S}=-e^2\phi'^2+\frac{p^2}{r^4}\,,\quad {\cal P}=-\frac{2ep\phi'}{r^2}\,.
\ee 
The $t$-component of the Maxwell equation \eqref{FE}, $(\nabla \cdot E)_t=0$ then yields
\be
\phi''+\phi'\frac{d}{dr}\lg\Bigl(\frac{4p^2{\cal L}_{{\cal P}^2}}{r^2}-r^2{\cal L}_{\cal S}\Bigr)=0\,.
\ee
In fact, without specifying NLE, one can integrate this equation once, to obtain
\be\label{phisol}
{\cal L}_{\cal S}=\frac{4p^2{\cal L}_{{\cal P}^2}}{r^4}+\frac{\beta}{r^2\phi'}\,,
\ee  
where $\beta$ is a dimensionless integration constant. Since ${\cal L}_{\cal S}$ and ${\cal L}_{{\cal P}^2}$ depend on $\phi$ (or more precisely its first derivatives) but not on $f$, this equation can be (at least in principle) integrated to obtain $\phi$. Once $\phi$ is known, the metric function $f$ can be obtained from the Einstein equation, say $H_{rr}=0$,
\be
f'+\frac{f}{r}+B(r)=0\,, 
\ee
where 
\be 
B(r)=4re^2\phi'^2 {\cal L}_{\cal S}-
2r(2{\cal P}^2{\cal L}_{{\cal P}^2}-{\cal L})-\frac{1}{r}\,, 
\ee
which yields a solution 
\be\label{fsol}
f=-\frac{\int B(r)rdr}{r}-\frac{2M}{r}\,, 
\ee
where $M$ is an integration constant.
The remaining equations are then automatically satisfied. We refer to \cite{Bokulic:2021dtz} for the discussion of thermodynamics of these solutions. 
}

\subsection{Newman--Janis algorithm}

{Starting from a static solution \eqref{sss}, there is a hope (fulfilled in the Maxwell/vacuum case) that one could obtain the corresponding rotating solution by the Newman--Janis algorithm, e.g. \cite{Newman:1965tw, Gurses:1975vu, Azreg-Ainou:2014pra, Toshmatov:2017zpr} (see also \cite{Erbin:2016lzq}). The `recipe' goes as follows. i) Start from a general spherical spacetime characterized by two metric functions $f=f(r)$ and $g=g(r)$, 
\be
ds^2=-f dt^2+\frac{dr}{f}+gd\Omega^2\,, 
\ee 
and proceed to the Eddington--Finkelstein coordinates, $(u,r,\theta,\varphi)$,
\be
du=dt-\frac{dr}{f}\,. 
\ee
The corresponding inverse metric can then be written as
\be\label{nullframe}
g^{\mu\nu}=-l^\mu n^\nu-l^\nu n^\mu+m^\mu {\bar m}^\nu+m^\nu {\bar m}^\mu\,, 
\ee
where the (complex) null frame reads
\be\label{nullor}
l=\partial_r\,,\quad n=\partial_u-\frac{f}{2}\partial_r\,,\quad m=\frac{1}{\sqrt{2g}}\Bigl(\partial_\theta+\frac{i}{\sin\theta}\Bigr)\partial_\phi\,. 
\ee
ii) Perform a complex coordinate transformation
\be\label{NJtransf}
u\to u-ia \cos\theta\,,\quad r\to r+ia\cos\theta\,, 
\ee
where $a$ is the rotation parameter, 
and replace $f\to F(r,a,\theta)$ and $g\to \Sigma(r,a,\theta)$. Of course, the transformation \eqref{NJtransf} affects the null frame \eqref{nullor}, as $\partial_\theta\to \partial_\theta+ia\sin\theta(\partial_u-\partial_r)$, that is, 
\ba
l&\to& \partial_r\,,\quad 
n\to \partial_u-\frac{F}{2}\partial_r\,,\nonumber\\
m&\to&  \frac{1}{\sqrt{2\Sigma}}\Bigl(\partial_\theta+ia\sin\theta(\partial_u-\partial_r)+\frac{i}{\sin\theta}\Bigr)\partial_\phi\,.  
\ea
Expression \eqref{nullframe} then defines the new metric by inversion. 
iii) Return back to the Boyer--Lindquist coordinates,
\be
du=dt+\lambda(r)dr\,,\quad d\varphi=d\varphi+\chi(r)dr\,, 
\ee
requiring that the only non-diagonal component of the metric is that of $g_{t\varphi}$. Together with imposing $g=r^2$, this fixes the above functions $F$ and $\Sigma$. The resulting metric then takes the following Carter's form \cite{Carter:1968ks}:  
\ba
 ds^2&=&-\frac{\Delta}{\Sigma}(dt-a\sin^2\!\theta d\varphi)^2+\frac{\Sigma}{\Delta} dr^2+\Sigma d\theta^2\nn\\
 &&+\frac{\sin^2\!\theta}{\Sigma}\Bigl[(r^2+a^2)d\varphi-adt\Bigr]^2\,,
\ea
where
\be
\Sigma=r^2+a^2\cos^2\!\theta\,,\quad \Delta=r^2f+a^2\,. 
\ee
Note that such a ``solution'' is completely characterized by a single metric function $f$ of the corresponding static solution. 
}

{
Of course, in the NLE case the generated metric  also has to be supplemented by the corresponding `rotating' vector potential $A$. For example, the following proposal: 
\ba
A&=&\frac{p\cos\theta}{\Sigma}\Bigl[(r^2+a^2)d\varphi-adt\Bigr]\,, 
\ea 
has been used in \cite{Toshmatov:2017zpr} to construct the rotating magnetically charged solutions. 
}

{
However, the above Newman--Janis generated rotating spacetime does not solve the corresponding NLE equations, e.g. \cite{Rodrigues:2017tfm}. In fact, as we shall see this is true even at the linear $O(a)$ level. To show this, consider a slowly rotating limit of the above metric and potential, obtaining thus: 
\ba\label{ansatz1}
ds^2&=&-fdt^2+\frac{dr^2}{f}+2ar^2\sin^2\!\theta h dtd\varphi+r^2d\Omega^2\,,\label{m1}\\
A&=&p\cos\theta\Bigl(d\varphi-\frac{a\omega}{r^2}dt\Bigr)\,,\label{m2} 
\ea
where, in the above we have $\omega=1$, and  
\be\label{hf}
h=\frac{f-1}{r^2}\,. 
\ee
As we shall see in the next section, the slowly rotating magnetically charged solutions of NLE can be obtained in the form \eqref{m1} and \eqref{m2}. However, the restriction \eqref{hf}, following from the {standard Newman--Janis} algorithm, is too strong and consistent only with the Maxwell theory (among all restricted theories \eqref{restrict}). The same conclusion about \eqref{hf} remains valid also in the electrically charged case.
In other words, the {standard Newman--Janis} algorithm fails to produce rotating solutions for any non-trivial NLE. This is similar to the recent observation \cite{Gray:2021roq} that the Einstein gravity is the only theory (up to quartic corrections in curvature) that admits the slowly rotating spacetimes of the form \eqref{m1} with $h$ given by \eqref{hf}. 
}

\section{Slowly rotating electric solutions}\label{Sec4a}
{
In this and next sections we shall show how (at least in principle) one can construct slowly rotating spacetimes coupled to any NLE. To this purpose, we shall consider the electrically and magnetically charged cases separately (and leave the more complicated dyonic case to the future studies). In both cases, we 
impose the generalized Lense--Thirring ansatz for the metric:
\be\label{metric}
ds^2=-fdt^2+\frac{dr^2}{f}+2ar^2\sin^2\!\theta h dtd\varphi+r^2d\Omega^2\,,
\ee
taking into account that at least at the $O(a)$ order, the condition $N=1$, \eqref{N1}, has to remain valid. In what follows, we shall consistently work to the linear order in the rotation parameter $a$.   
}

\subsection{Finding electric solutions}
{
For the electric solutions, we choose the following ansatz for the vector potential:
\be\label{ansatzEl}
A=e\phi(dt-a\omega \sin^2\!\theta d\varphi)\,, 
\ee
where $\phi$ corresponds to the static solution and $\omega=\omega(r)$ captures the effect of rotation. In this case we find 
\be\label{elinvariants}
{\cal S}=-e^2\phi'^2+O(a^2)\,,\quad 
{\cal P}=-\frac{4e^2\phi\phi'\omega\cos\theta}{r^2}a+O(a^3)\,. 
\ee
Again, since ${\cal P}$ is linear in $a$, and ${\cal L}={\cal L}({\cal S},{\cal P}^2)$, then at a given $O(a)$ order $T_{\mu\nu}$ takes the following simplified form: 
\be\label{Tmunusimpl}
4\pi T_{\mu\nu}=f^0F_{\mu\sigma}F_{\nu}{}^\sigma+g^0 g_{\mu\nu}\,, 
\ee
where 
\be \label{f0simpl}
f^0=2 {\cal L}_{\cal S}|_{{\cal P}=0}\,,\quad g^0=-{\cal L}|_{{\cal P}=0}\,,
\ee
both being functions of $\phi'$ (independent of $\omega$ and $f$).
}

{
Considering Einstein equations $H_{rr}=0=H_{\varphi\varphi}$, and solving them algebraically for $f^0$ and $g^0$, we obtain: 
\be\label{f0g0el}
f^0=\frac{\zeta}{4e^2r^2\phi'^2}\,,\quad 
g^0=\frac{rf''+2f'}{4r}\,, 
\ee
where 
\be\label{zeta}
\zeta=r^2f''-2f+2\,. 
\ee
At the same time $(\nabla \cdot E)_t=0$ can algebraically be solved for ${\cal L}_{\cal S}'$, and yields
\be\label{LSel}
{\cal L}_{\cal S}'=-\frac{{\cal L}_{\cal S}(r\phi''+2\phi')}{r\phi'}\,.
\ee
The remaining non-trivial equations are then 
$H_{t\varphi}=0$ and $(\nabla \cdot E)_\varphi=0$ (both being of the order $a$). They explicitly give: 
\ba
r^4\phi'h''+4r^3\phi'h'-
\phi\zeta\omega'-\phi'\zeta \omega&=&0\,,\label{46}\\
A\omega''+B\omega'+C\omega-r^4\phi'^2{\cal L}_{\cal S}h'&=&0\,, \label{56}
\ea
where 
\ba
A&=&r^2\phi\phi'f {\cal L}_{\cal S}\,,\nonumber\\
B&=&-r{\cal L}_{\cal S}\Bigl(rf\phi\phi''-r\phi\phi'f'-2rf\phi'^2+2f\phi\phi'\Bigr)\,,\\
C&=&\phi'\Bigl(-8e^2\phi\phi'^2{\cal L}_{{\cal P}^2}+r\phi'{\cal L}_{\cal S}(rf'-2f)-2\phi{\cal L}_{\cal S}\Bigr)\,,\nonumber 
\ea
and ${\cal L}_{\cal S}$ and ${\cal L}_{{\cal P}^2}$ are expressed at ${\cal P}=0$, that is, they are functions of $\phi'$ but not $\omega$.
}

{Thus, we have a simple procedure for determining the slowly rotating electrically charged solutions. The functions $f$ and $\phi$ are those of the corresponding static solution, given by \eqref{fsol} and \eqref{phisol}, after setting $p=0$. Eqs.~\eqref{46} and \eqref{56} then represent coupled ordinary differential equations for `rotating' functions $\omega$ and $h$. Obviously, one can express $h'$ from the second equation, and by plugging this back to the first one, obtain a 3rd-order ODE for $\omega$. As discussed in conclusions, such an equation is guaranteed to have a `nice' solution for $\omega$, which then yields $h$ by integrating \eqref{56}. 
}

\subsection{Maxwell uniqueness}

{
Let us now impose the conditions \eqref{h}, $h=(f-1)/r^2$, upon which Eq.~\eqref{46} gives
\be 
\Bigl[\phi(\omega-1)\Bigr]'\,\zeta=0\,.
\ee
In other words, we find that for any non-trivial NLE, we have to have 
\be
\omega=1+\frac{c}{\phi}\,, 
\ee
for some (dimensionful) constant  $c$. 
Plugging this into Eq.~\eqref{56} then yields 
\be\label{conel1}
r{\cal L}_{\cal S}\phi'+{\cal L}_{\cal S}(\phi+c)+4e^2\phi'^2{\cal L}_{{\cal P}^2}(\phi+c)=0\,.
\ee
For restricted class of theories, \eqref{restrict}, we have ${\cal L}_{{\cal P}^2}=0$,\footnote{More generally, the theory ${\cal L}({\cal S},{\cal P}^2)$ is admissible provided the solution of \eqref{conel1} is consistent with the solution of \eqref{phisol} (with $p=0$). This requirement seems rather restrictive. In particular, we have checked that it is not satisfied for the ModMax theory \cite{Bandos:2020jsw, Kosyakov:2020wxv} -- this theory thus does not admit slowly rotating electric solutions with $h=(f-1)/r^2$. On the other hand, a theory given by ${\cal L}=({\cal S}^4+{\cal P}^4)^{1/4}$ admits trivially Maxwell-like solutions. } and the latter equation  can be integrated to give
\be
\phi=\frac{1}{r}-c\,. 
\ee
It is now obvious that the constant $c$ is unphysical and only corresponds to the gauge for the vector potential -- it can be gauged away by  $A\to A+d\lambda$ where $\lambda=ec t$. Thus without loss of generality, we have 
\be
\omega=1\,,\quad \phi=\frac{1}{r}\,. 
\ee 
Eq. \eqref{phisol} (with $p=0$) then yields the Maxwell theory. 
Thus we have proved the following:\\
{\bf Theorem.} {\em For restricted class of theories, \eqref{restrict}, the only NLE consistent with $h=(f-1)/r^2$ for the ansatz \eqref{metric} and \eqref{ansatzEl} is the Maxwell theory.}\\
In particular, this implies:\\
{\bf Collorary.} {\em Electrically charged spacetimes generated by the {standard Newman--Janis} algorithm do not solve the corresponding NLE equations following from \eqref{restrict}, not even at the linear $O(a)$ level. }  \\
}

\subsection{Special NLE with $\omega=1$}

{In the above we have established that imposing $h=(f-1)/r^2$ leads to the Maxwell theory, and in particular  implies that one has to have $\omega=1$. Let us now ask the `opposite': imposing   
\be\label{omega1}
\omega=1\,, 
\ee
can we have any non-trivial NLE? The partial motivation to study this question stems from the NLE literature, e.g. \cite{CiriloLombardo:2005qmc, Hendi:2010bk}, where the assumption \eqref{omega1} is automatically assumed.  As we shall now show, apart from the Maxwell theory,  there is yet another special NLE consistent with \eqref{omega1}, given by \eqref{new} below. This theory is, however, distinct from the NLE theories studied in \cite{CiriloLombardo:2005qmc, Hendi:2010bk}, invalidating thus some of the the results in these papers. This theory also provides an example of NLE where slowly rotating electric solutions can be explicitly constructed. 
}

{
To construct our special NLE,  let us return to Eq.~\eqref{46} and impose \eqref{omega1}. In this case, this equation can be integrated for $h$, and gives 
\be
h=\frac{f-1}{r^2}-\frac{2M_0}{r^3}+h_0\,, 
\ee
where $M_0$ and $h_0$ are the integration constants. Here $h_0$ can be reabsorbed by redefining $\varphi$, namely, $d\varphi\to d\varphi -ah_0dt$. In other words, $h_0$ is not physical and we can set $h_0=0$. On the other hand $M_0$ seems physical as it `redefines' the asymptotic angular momentum. Of course, one possibility is to consider $M_0=0$, in which case we are back to the Maxwell case. On the other hand, considering $M_0$ non-trivial, Eq.~\ref{56} then yields 
\be
4e^2{\cal L}_{{\cal P}^2}\phi\phi'^2+{\cal L}_{\cal S}(r+3M_0)\phi'+\phi{\cal L}_{\cal S}=0\,. 
\ee
Focusing on the restricted theories \eqref{restrict}, the latter can be integrated and gives 
\be
\phi=\frac{1}{r+3M_0}\,. 
\ee 
The corresponding metric function $f$ is then obtained by integrating \eqref{phisol} where the l.h.s. is given by one half of the first expression in \eqref{f0g0el}. This then yields  
\ba
f&=&1+\frac{4\beta e^2}{9M_0^2}-\frac{2M+8\beta e^2/(9M_0)}{r}-\frac{8r\beta e^2}{27M_0^3}\nonumber\\
&&+r^2\Bigl(\Lambda-\frac{8\beta e^2}{81M_0^4}\lg\bigl(\frac{r}{r+3M_0}\bigr)\Bigr)\,, 
\ea
where $M$ and $\Lambda$ are the integration constants. Interestingly, for large $r$ this has the following expansion:
\be
f\approx 1-\frac{2M}{r}+\Lambda r^2-\frac{2\beta e^2}{r^2}+\frac{24\beta e^2 M_0}{5 r^3}+O\Bigl(\frac{1}{r^4}\Bigr)\,, 
\ee
which upon setting $\Lambda=0$ and $\beta=-1/2$ has the required Reissner--Nordstrom asymptotic behavior. For small enough positive $M_0$, we have (up to) two horizons, shielding singularity at $r=0$. Note also that the electromagnetic field is regular on the horizon, by the token of \eqref{elinvariants}.

The corresponding theory can easily be constructed from \eqref{phisol}. Namely, we have 
\be \label{LSco}
{\cal L}_{\cal S}=-\frac{1}{2r^2\phi'}=
\frac{(r+3M_0)^2}{2r^2}=\frac{1}{2}\bigl(1-s\bigr)^{-2}\,,
\ee   
where 
\be
s=\Bigl(-\frac{{\cal S}}{{\cal S}_0}\Bigr)^\frac{1}{4}\,,\quad  
{\cal S}_0=\frac{e^2}{(3M_0)^4}\,.
\ee
Eq.~\eqref{LSco} can be integrated to yield
\be\label{new}
{\cal L}=2{\cal S}_0\Bigl(\frac{s^3+3s^2-4s-2}{2(1-s)}-3\lg(1-s)+1\Bigr)\,. 
\ee
which obeys \eqref{POC}.
Of course, in here ${\cal S}_0$ is the fundamental coupling constant, that gives rise to the modification related to $M_0$ above. 
}

{
To conclude, among restricted NLE theories \eqref{restrict}, there are two theories that yield the Lense--Thirring solutions with $\omega=1$: the Maxwell theory and the theory defined by the Lagrangian \eqref{new}. Surprisingly, this Lagrangian is identical to the one obtained in \cite{Tahamtan-PRD:2021} as the only NLE model admitting electromagnetic radiation in Robinson--Trautman class of spacetimes.
}

\section{Slowly rotating magnetic solutions}\label{Sec4b}

\subsection{Finding magnetic solutions}

{To find the magnetic solutions, we supplement the metric \eqref{metric} with the following ansatz for the vector potential: 
\be\label{Amag}
A=p\,\cos\theta\left(d\varphi-\frac{a\,\omega}{r^2} dt\right)\,, 
\ee
where $\omega=\omega(r)$ is a new vector potential function. 
The field invariants \eqref{SP} now read
\be
{\cal S}=\frac{p^2}{r^4}+O(a^2)\,,\quad 
{\cal P}=\frac{2p^2\cos\theta(r\omega'-2\omega)}{r^5}a+O(a^3)\,. 
\ee
Note that since ${\cal P}$ is linear in $a$, and ${\cal L}={\cal L}({\cal S},{\cal P}^2)$, then at a given $O(a)$ order $T_{\mu\nu}$ takes a simplified form: 
\be\label{Tmunusimpl}
4\pi T_{\mu\nu}=f^0F_{\mu\sigma}F_{\nu}{}^\sigma+g^0 g_{\mu\nu}\,, 
\ee
where 
\be \label{f0simpl}
f^0=2 {\cal L}_{\cal S}|_{{\cal P}=0}\,,\quad g^0=-{\cal L}|_{{\cal P}=0}\,,
\ee
both being explicit functions of $r$ (independent of $\omega$ and $f$).
Eq.~\eqref{fsol} with $e=0$ immediately gives 
\be\label{fmag}
f=1-\frac{2M}{r}-\frac{2\int r^2{\cal L}(r)dr}{r}\,. 
\ee
}

{
Solving algebraically 
$H_{rr}=0=H_{\varphi\varphi}$ for $f^0, g^0$ 
yields 
\be\label{f0g0mag}
f^0=\frac{r^2\zeta}{4p^2}\,,\quad 
g^0=\frac{1}{2r^2}(rf'+f-1)\,, 
\ee
where $\zeta$ is given by \eqref{zeta}, $\zeta=r^2\,f''-2f+2$.
Eliminating further 
$\frac{d}{dr}{\cal L}_{\cal S}$ from 
$(\nabla \cdot E)_t=0$, and plugging these to $H_{t\varphi}=0$ and $(\nabla\cdot E)_t=0$, gives the following two $O(a)$ equations: 
\ba
r^4\,f\,h''+4\,r^3\,f\,h'-r^2\,\zeta\,h-\zeta\,\omega&=&0\,,\label{35}\\ 
A\omega''+
B\omega'+C\omega-2r^6{\cal L}_{\cal S}h&=&0\,,\label{355}
\ea
where
\ba
A&=&(r^6{\cal L}_{\cal S}-4p^2r^2{\cal L}_{{\cal P}^2})f\,, \nonumber\\
B&=&rf(r^5{\cal L}_{\cal S}'-2r^4{\cal L}_{\cal S}-4p^2r{\cal L}_{{\cal P}^2}'+24p^2{\cal L}_{{\cal P}^2})\,,\\
C&=&f(8p^2r {\cal L}_{{\cal P}^2}'-2r^5{\cal L}_{\cal S}'+2r^4{\cal L}_{\cal S}-40p^2{\cal L}_{{\cal P}^2})-2r^4 {\cal L}_{\cal S}\,,\nonumber
\ea
Eqs.~\eqref{35} and \eqref{355}  represent, two coupled ordinary differential equations for $\omega$ and $h$. 
While these equations can be easily decoupled -- they result in higher(4th)-order linear ODEs with variable coefficients. (We shall briefly comment on finding the corresponding solutions in conclusions.)  This procedure is illustrated in Appendix~\ref{AppB} where we construct slowly rotating magnetized black holes in the `Square Root' NLE.  
}

\subsection{Maxwell uniqueness}

{
Let us now impose \eqref{h},  $h=(f-1)/r^2$. Then Eq.~\eqref{35} immediately yields 
\be
(\omega-1)\zeta=0\,, 
\ee
and for any non-trivial NLE, we must have 
\be
\omega=1\,. 
\ee
Returning back to $(\nabla \cdot E)_t=0$ then yields that 
\be\label{hardtosolve}
r^5{\cal L}_{\cal S}'-4p^2r\frac{d}{dr}{\cal L}_{{\cal P}^2}+20p^2{\cal L}_{{\cal P}^2}=0\,. 
\ee
Obviously, for the restricted class of theories \eqref{restrict} we have just proved that one has to have ${\cal L}_{\cal S}=const.$, which is only consistent with the Maxwell theory.\footnote{Again, Eq.~\eqref{hardtosolve} is not satisfied for the ModMax theory, and trivially works for ${\cal L}=({\cal S}^4+{\cal P}^4)^{1/4}$.} We have thus proved the following:\\
}
{
{\bf Theorem.} {\em Among all restricted NLE theories \eqref{restrict}, Maxwell theory is the only one that admits the magnetically charged slowly rotating solutions of the form \eqref{metric}, \eqref{Amag} with the restriction $h=(f-1)/r^2$. In particular, this means that the {standard Newman--Janis} algorithm fails to produce solutions already at the linear $O(a)$ level.  
}}

This theorem, in particular, invalidates `solutions' constructed  in  \cite{Toshmatov:2014nya, Toshmatov:2017zpr}.


\section{Conclusion}\label{Sec5}
{
In this paper we have  
analysed slowly rotating generalizations of static spacetimes sourced by NLE. 
To this end we have presented a generalized Lense--Thirring ansatz for the metric \eqref{metric} together with a simple ansatz for the vector potential in electric \eqref{ansatzEl} and magnetic \eqref{Amag} cases and shown that with these one can solve (at least in principle) the corresponding Einstein-NLE equations to the linear order in the rotation parameter. Dyonic case seems more complicated and we leave it for the future studies. {To illustrate this procedure, we have found two explicit examples  where the corresponding solutions can be found in a closed form. (The detailed analysis of these solutions is left for future studies.)}  

We {have also proved the ``No Go Theorem'' which establishes} that the Maxwell theory is the only NLE among all theories \eqref{restrict} that admits function $h$ given by the `natural' expression \eqref{h}.  {This in particular shows that the standard Newman--Janis algorithm (which leads to the form \eqref{h} in the slow rotation approximation)} fails to produce rotating solutions in NLE even at the lowest (linear) level in rotation parameter $a$, as well as shows that a number of (slowly rotating) metrics constructed in previous studies cannot satisfy the corresponding equations of motion. We have also pointed out that i) some of the previous attempts to construct rotating solutions in NLE  used too simplistic $(\omega=1)$ ansatz for the vector potential, while ii) some other previously constructed solutions actually do not present slowly rotating black holes but rather correspond to weakly NUT-charged solutions (see Appendix~\ref{AppA} {where these solutions are constructed for the general case}). {This effectively leaves no solutions at all for (slowly) rotating black holes in NLE (see, however, the recent progress in \cite{Diaz:2022roz, Ayon-Beato:2022dwg}).  
}

{Our work opens several new directions for future studies. 
First, the above No Go Theorem strictly speaking only regards the restricted NLEs \eqref{restrict}. It is possible that once a more general setting of  \eqref{general} is considered, there are other theories which allow for \eqref{h}, see Eq.~\ref{conel1} and the corresponding discussion in the footnote. It would be interesting to construct examples of such non-trivial theories. }

{
Second, while we have shown that the standard Newman--Janis algorithm does not give raise to the corresponding rotating solutions, we cannot dismiss a possibility that an appropriately {\em modified 
Newman--Janis} algorithm cannot be formulated for NLE theories. The explicit slowly rotating solutions found in this paper may provide a test ground for finding such an algorithm.}

{Third, during our investigation, we have stumbled upon a {\em special} example of NLE, whose slowly rotating electric solutions are distinguished by the `Maxwell-like' $(\omega=1)$ form of the 
gauge potential. Interestingly, this is the same theory discovered recently in \cite{Tahamtan-PRD:2021} as the only theory of NLE admitting radiation in the Robinson--Trautman class of spacetimes. This theory certainly deserves further attention in the future.}  

Fourth, as we have seen above, the general solution for the slowly rotating NLE hinges on solving a higher order linear ODE with variable coefficients. 
Assuming the coefficients of the equation are continuous functions, the general theory of ODE guarantees existence and uniqueness of solution provided some initial conditions are prescribed. Furthermore, the solution of such $n$-th order ODE would be at least $C^{n-1}$.
In this way the procedures detailed above provide solution to our stated problem.
However, one would like to have more then just existence result. To this end one can transform the higher order ODE into first-order system and employ some of the approximate solution methods. One such method is the Magnus expansion \cite{Magnus:1954zz} which gives the solution as an exponential of a series containing integrals of nested commutators of the coefficient matrix. The convergence is controlled by suitable norm of the coefficient matrix and even truncated solutions often capture the main features of the complete solution.  

{Historically, it took almost fifty years to upgrade the slowly rotating charged solutions of Lense--Thirring to the full non-linear Kerr--Newmann geometry. It will be interesting to see if some of the hereby presented slowly rotating solutions can be promoted to full (possibly analytic) charged and rotating solutions in some non-trivial NLE. 
}

\section*{Acknowledgements}
We would like to thank the anonymous referee for helping us to improve our manuscript.  
D.K. acknowledges the support from the Perimeter Institute for Theoretical Physics and the
Natural Sciences and Engineering Research Council of Canada (NSERC).
T.T was supported by Research Grant No. GA\v{C}R 21-11268S and O.S by Research Grant No. GA\v{C}R 22-14791S.
Research at Perimeter Institute is supported in part by the Government of Canada through the Department of Innovation, Science and Economic Development Canada and by the Province of Ontario through the Ministry of Colleges and Universities. 
Perimeter Institute and the University of Waterloo are situated on the Haldimand Tract, land that was promised to the Haudenosaunee of the Six Nations of the Grand River, and is within the territory of the Neutral, Anishnawbe, and Haudenosaunee peoples.

\appendix

\section{Magnetic solutions in Square Root electrodynamics}\label{AppB}

{
In this appendix we shall construct slowly rotating magnetized solution in the so called {\em Square Root} model of NLE, characterized by 
\be
{\cal L}=-\beta \sqrt{S}\,, 
\ee 
where $\beta$ is a dimensionful coupling constant, with dimensions $1/L$.
This Lagrangian represents a strong field regime of many models of NLE, the Born--Infeld theory for example. It was 
originally proposed by Nielsen and Olesen \cite{Nielsen1973} to treat the so-called dual string in flat spacetime.
It also gives rise \cite{Guendelman2006} to the confinement potential \cite{tHooft:2002pmx}, see also \cite{Guendelman2014} for recent developments on non-linear gauge theories containing "Square Root" Lagrangians. 
Of course, the model can also be generalized to curved spacetime, and was recently discussed in \cite{ Tahamtan-PRD:2020, Tahamtan-RAG19,Tahamtan-Kundt:2017}.  
Note that when considering only the magnetic field all energy conditions are satisfied unlike the case of pure radial electric field. 
}

{To construct the magnetized solution, we adopt the ansatz \eqref{metric} together with the potential \eqref{Amag}, and follow the procedure outlined in Sec.~\ref{Sec4b}. 
Namely, the static metric function $f$, \eqref{fmag}, is given by
\be\label{f-squareroot}
 f=1-\frac{2M}{r}+2\beta p\,.
\ee
Note that this modification of the Schwarzschild solution is related to the solid angle deficit/excess (depending on the sign of $\beta p$). Such solution also represents the geometry outside the core of the so-called global monopole, a spacetime defect created by a gravitating triplet of scalar fields whose original $O(3)$ symmetry is spontaneously broken to $U(1)$. Global monopole was extensively discussed in literature, see, e.g. \cite{Letelier-1979, Barriola, global1} for original works and some more recent work by two of the authors \cite{Tahamtan-quantum:2014, Tahamtan-Boosting}.
}

{For $M=0$, Eqs. \eqref{35} and \eqref{355} yield the following solution for $h$ and $\omega$:
\ba\label{s11}
\omega&=&\frac{\omega_1}{r}+\omega_2 r^2+\omega_3r^{(\frac{1}{2}+q)}+\omega_4 r^{(\frac{1}{2}-q)}\,,\nonumber\\
h&=&-\frac{\omega_1}{r^3}-\omega_2-2\omega_3\beta p r^{(q-\frac{3}{2})}-2\omega_4\beta p r^{(-q-\frac{3}{2})}\,,\qquad
\ea
where $\omega_i$'s are the integration constants, and
\be
q=\frac{\sqrt{4p^2\beta^2+36 p\beta+17}}{4 p\beta+2}\,.
\ee
Obviously, we can eliminate $\omega_2$ by redefining $\varphi$, and so we set $\omega_2=0$. Writing $\omega_1=2M_0$ and setting $\omega_3=0=\omega_4$ for simplicity (though it would be interesting to study the physical meaning of these terms), we thus recover the following simple solution:
\be\label{sol_simple}
h=-\frac{2M_0}{r^3}\,,\quad \omega=\frac{2M_0}{r}\,. 
\ee
}

{When $M\neq 0$, the terms with `strange powers' of $r$ in \eqref{s11} are replaced by hypergeometric functions. However, setting again $\omega_3=0=\omega_4$, the solution \eqref{sol_simple} remains valid also in this case.}

\section{Taub-NUT solutions in NLE}\label{AppA}
Lorentzian Taub-NUT spacetimes \cite{taub1951empty, newman1963empty} represent an interesting class of axisymmetric (electro-vacuum) solutions of Einstein equations. Such solutions are characterized by the appearance of the so called Misner strings \cite{misner1963flatter} -- the singular rotating sources of angular momentum \cite{bonnor1969new}. As these strings extend all the way to infinity, the Taub-NUT  solutions are not asymptotically flat. They also feature various pathologies, such as the existence of closed timelike curves in the vicinity of Misner strings. As we shall see below, some of these solutions were in the NLE literature confused with the slowly rotating black hole solutions. 
To demonstrate this, we shall study the charged Taub-NUT solutions coupled to a general NLE \eqref{general}.

Namely, we seek the charged Taub-NUT solution in the following form:  
\ba
    ds^2 &=&{ -f\big(dt + 2n \cos\theta d \varphi\big)^2}+\frac{dr^2}{f} \nonumber\\
 &&\quad + (r^2+n^2)\left(d\theta^2 +\sin^2\!\theta d\varphi^2\right)\,\nonumber\\
A &=& \phi(dt + 2n\cos\theta d\varphi)\,, \label{TN}
\ea
where we  have denoted the NUT parameter by $n$, choosing a  symmetric distribution for Misner strings (which are located on both the north-pole and south-pole axes). The solution is characterized 
by a single metric function $f=f(r)$, and single gauge potential function $\phi=\phi(r)$.  
 
Using this ansatz, we find the following expressions for the invariant ${\cal S}$ and ${\cal P}$:
\be
{\cal S}=-\phi'^2+\frac{4n^2\phi^2}{(n^2+r^2)^2}\,,\quad {\cal P}=-\frac{4n\phi\phi'}{n^2+r^2}\,.
\ee
The $t$-component of the generalized Maxwell equation \eqref{FE} then yields the following ODE:
\ba
\phi''+\phi'\frac{d}{dr}\lg\Bigl(-(n^2+r^2){\cal L}_{\cal S}\Bigr)\qquad \qquad\qquad&&\nonumber\\
+\frac{2n\phi}{(n^2+r^2)^2}\Bigl(\frac{(n^2+r^2){\cal L}_{\cal P}'}{{\cal L}_{\cal S}}+2n\Bigr)&=&0\,. 
\ea
Since ${\cal L}_{\cal S}$ and ${\cal L}_{\cal P}$ depend only on $\phi$ but not $f$, this equation can (at least in principle) be integrated to yield solution for $\phi$. The $H_{rr}=0$ Einstein equation then yields a first-order equation for $f$:
{
\ba
f'+\frac{r^2-n^2}{r(n^2+r^2)}f
-\frac{1}{r}\Bigl(1-8n{\cal L}_{\cal P}\phi\phi'
\qquad  &&\nonumber\\
-2(n^2+r^2)({\cal L}+2{\cal L}_{\cal S}\phi'^2)\Bigr)&=&0\,, 
\ea
}
and the remaining equations are automatically satisfied. The explicit examples were constructed e.g. for the Born--Infeld theory \cite{garcia1984type} (see also \cite{breton2015nut}) or the recently constructed ModMax theory \cite{BallonBordo:2020jtw}.

In particular, let us consider a small $n$ expansion, expanding the metric and the gauge potential to the linear order in $n$. In this case the {\em weakly NUT-charged} solution is fully characterized by the static metric function $f$ and static electric potential of the corresponding NLE, determined from   
{
\ba
\phi''+\phi'\frac{d}{dr}\lg(-r^2{\cal L}_{\cal S})&=&0\,,\nonumber\\
f'+\frac{f}{r}-\frac{1}{r}\Bigl(1-2r^2({\cal L}+2 {\cal L}_{\cal S}\phi'^2)\Bigr)&=&0\,,
\ea
c.f. Eqs.~\eqref{phisol} and \eqref{fsol}.
(As always, here we assumed that ${\cal L}={\cal L}({\cal S},{\cal P}^2)$ and so ${\cal L}_{\cal P}\sim O(n)$.)}
In particular, for a specific NLE, the corresponding solutions were constructed in \cite{Hendi:2014xia, Lammerzahl:2018zvb} and confused with the slowly rotating black holes.


\providecommand{\href}[2]{#2}\begingroup\raggedright\endgroup

\end{document}